\documentclass[twocolumn,showpacs,showkeys,preprintnumbers,amsmath,amssymb,floatfix]{revtex4}

\usepackage{graphicx}
\usepackage{dcolumn}
\usepackage{bm}
\usepackage{epstopdf} 

\begin{document}

\title{Wall slip, shear banding, and instability\\in the flow of a triblock copolymer micellar solution}

\author{S\'{e}bastien Manneville}
\email{sebastien.manneville@ens-lyon.fr}
\affiliation{
Centre de Recherche Paul Pascal, UPR8641, \\
115 avenue Schweitzer, 33600 Pessac, FRANCE\\
Present address: Laboratoire de Physique - ENS Lyon, UMR5672\\
46 all\'ee d'Italie, 69364 Lyon cedex 07, FRANCE
}

\author{Annie Colin}
\affiliation{
Laboratoire Du Futur, UMR CNRS-Rhodia FRE2771,\\
178 avenue Schweitzer, 33607 Pessac, FRANCE
}

\author{Gilles Waton and Fran\c{c}ois Schosseler}
\email{schossel@ics.u-strasbg.fr}
\affiliation{
Institut Charles Sadron, UPR22\\
6 rue Boussingault, 67083 Strasbourg cedex, FRANCE
}

\date{\today}

\begin{abstract}
The shear flow of a triblock copolymer micellar solution (PEO--PPO--PEO Pluronic P84 in brine) is investigated using simultaneous rheological and velocity profile measurements in the concentric cylinder geometry. We focus on two different temperatures below and above the transition temperature $T_c$ which was previously associated with the apparition of a stress plateau in the flow curve. (i)~At $T=37.0^\circ$C~$<T_c$, the bulk flow remains homogeneous and Newtonian-like, although significant wall slip is measured at the rotor that can be linked to an inflexion point in the flow curve. (ii)~At $T=39.4^\circ$C~$>T_c$, the stress plateau is shown to correspond to stationary shear-banded states characterized by two high shear rate bands close to the walls and a very weakly sheared central band, together with large slip velocities at the rotor. In both cases, the high shear branch of the flow curve is characterized by flow instability. Interpretations of wall slip, three-band structure, and instability are proposed in light of recent theoretical models and experiments.
\end{abstract}


\maketitle

\section{\label{sec:level1}Introduction}

Triblock copolymers PEO--PPO--PEO based on poly(ethylene oxide) (PEO) and poly(propylene oxide) (PPO) have interesting self-assembly behavior in aqueous solutions due to their amphiphilic character \cite{wanka90,alex94,kozlov00}. PPO is soluble in water only at low concentrations and for  molecular weights smaller than about 4000 \cite{armstrong95}. As a homopolymer, PEO is insoluble in water at low and high temperatures. The binodal appears as a closed loop in the temperature--concentration phase diagram, and the miscibility region shrinks with increasing molecular weight \cite{saeki76,bae91,bae93}. Thus the hydrophilic/hydrophobic balance of these copolymers can be tuned by changing the PEO/PPO molar ratio and the temperature. Above the critical micelle concentration (cmc) and the critical micellar temperature (cmt), the copolymer chains aggregate and form micelles with a PPO core and a PEO corona \cite{mortensen01}. The cmc and the cmt both depend on the total molecular weight \cite{wanka90,alex94,kozlov00} and on the concentration of added components like, e.g., monovalent salts \cite{almgren91,bahadur92,alex97}.

The structure of the micelles has been elucidated by using scattering techniques \cite{mortensen01,pedersen02}. Above the cmt and the cmc, spherical micelles are obtained with the aggregation number being an increasing function of the temperature \cite{brown91,mortensen93a}. At low temperature many water molecules are still present in the PPO core but are progressively expelled as the temperature increases. Yet the radius of the core increases and there is a growing entropic contribution from the stretched PPO blocks to the free energy per micelle. Hence the aggregation number cannot increase indefinitely and a spheroidal or cylindrical shape can be expected \cite{brown91,mortensen93a} if the size of the PEO blocks is not too large compared to that of the PPO block \cite{linse93}. This is basically the same packing parameter criterion as the one used for conventional surfactants \cite{isra91}. Clear experimental evidence for the growth of elongated micelles has been provided by scattering techniques for the commercial Pluronic copolymers P85 ($M_{w}=4600$~g~mol$^{-1}$, PEO weight fraction $\approx 0.5$) \cite{mortensen93b,schillen94,jorgensen97} and P84 ($M_{w}=4200$~g~mol$^{-1}$, PEO weight fraction $\approx 0.4$) \cite{liu98,michels01,aswal01,duval05}.

At low concentration of the copolymer micelles, the temperature-induced micellar growth has a huge influence on the viscoelastic properties. Typically, for a Pluronic P84 weight fraction of 0.04 in 2~M NaCl brine, the zero-shear viscosity increases by a factor 10$^{5}$ when the temperature increases from 30$^\circ$C to 40$^\circ$C \cite{waton04}. Below 30$^\circ$C the solution contains only dilute spherical micelles with a negligible effect on the viscosity. As the aggregates grow into wormlike micelles, the solution enters the semidilute regime of living entangled polymers and the zero-shear viscosity jumps accordingly. Above about 41$^\circ$C, the solvent becomes progressively poorer for the PEO corona and the aggregates become more compact as the cloud point is approached \cite{duval05}.

The non-Newtonian behavior under steady shear is also strongly influenced by micellar growth. The steady-state flow curves (shear stress $\sigma$ vs\ shear rate $\dot\gamma$) measured in the concentric cylinder geometry display a different behavior below and above a transition temperature $T_c\approx 37.7^\circ$C at weight fraction 0.04 \cite{waton04}. Below this temperature the shear stress is a monotonic increasing function of the imposed shear rate, while above $T_c$ the flow curve exhibits two increasing branches separated by a stress plateau \cite{waton04}. The presence of this plateau is often interpreted as the hallmark of shear-banded flows, i.e., of the existence of a non-uniform velocity gradient across the gap of the rheometer. The occurrence and the stability of non-uniform shear flows have been studied theoretically for about twenty years \cite{mcleish86, spenley93, schmitt95, spenley96, olmsted97, yuan99, olmsted99, goveas99, radu00, olmsted00, lu00, goveas01, yuan02, bautista02, fielding03, fielding03b, fielding03c, fielding04, cook04, aradian05, mikh05, fielding05, yesilata06,rossi06}. The existence of a plateau in the flow curve has been indeed observed in a number of experimental systems, mainly in conventional surfactant solutions \cite{rehage91, khatory93, berret94, mair96, britton97, bolten97, koch98, britton99, fischer00, fischer02, escal03, salmon03, mendez03, radu03, lerouge04, lopez04, becu04b, herle05, lee05, hu05, azzouzi05,vdg06}. However only a few studies have been able to measure directly the velocity profile in the rheometer gap using NMR velocimetry \cite{mair96, britton97, britton99, fischer00, lopez04}, particle image velocimetry \cite{koch98, hu05}, heterodyne dynamic light scattering \cite{salmon03,vdg06} or ultrasonic speckle velocimetry (USV) \cite{becu04b}. The recently developped USV technique allows one to measure velocity profiles with a temporal resolution of 0.02 to 2~s depending on the shear rate \cite{manneville04}. Thus the stability of inhomogeneous velocity profiles can be investigated and the transition to unstable flow regimes can be studied.

In Ref.~\cite{waton04} two of us used small-angle light scattering to study Pluronic P84 solutions in brine sheared in the concentric cylinder geometry. Strong indications for shear-induced demixing below $T_c$ and for shear banding above $T_c$ were found but no direct evidence from velocity measurements was provided. The aim of the present paper is to use the USV technique to simultaneously measure the rheology and velocity profiles of the same solutions in experimental conditions very close to those of Ref.~\cite{waton04}. We study the system at different temperatures, below and above the transition temperature $T_c$ (where a plateau appears in the flow curve). We are thus able to follow the evolution of the velocity profiles as a function of the applied shear rate and of the temperature, and to identify different flow regimes that turn out to be more complex than those usually observed in shear-banded flows.

The paper is organized as follows. Experimental details about the samples and about the rheological and velocimetry setups are given in Sect.~\ref{sec:exp}. Results obtained at two different temperatures are described in Sect.~\ref{sec:res}. We show that (i) at $T=37.0^\circ$C the bulk flow remains homogeneous and Newtonian-like, although significant wall slip is measured at the rotor that can be linked to an inflexion point in the flow curve, and (ii) at $T=39.4^\circ$C the stress plateau in the flow curve is associated to stationary shear-banded states characterized by two high shear rate bands close to the walls and a very weakly sheared central band, together with large slip velocities at the rotor. In both cases, on the high shear branch of the flow curve, the flow becomes highly non-stationary and two-dimensional or even possibly three-dimensional. These results are further discussed in Sect.~\ref{sec:discuss} where interpretations of wall slip, three-band structure, and instability are proposed in light of recent theoretical models and experimental results on shear banding.

\section{Experimental section}
\label{sec:exp}

\subsection{Sample preparation}

We used without further purification the commercial Pluronic P84 supplied by BASF. The total molecular weight is 4200~g~mol$^{-1}$ and the nominal stoechiometry is PEO$_{19}$--PPO$_{43}$--PEO$_{19}$. With 2~M added NaCl,  microcalorimetric measurements \cite{michels01} show that, for a 0.04 weight fraction, the cmt and the cloud point are at about 2$^\circ$C and 43$^\circ$C, respectively. The onset of micellar growth is about 30$^\circ$C \cite{michels01,duval05}. The transition to the semidilute regime of wormlike micelles is estimated to occur at about 35$^\circ$C, when the zero-shear viscosity has increased by a factor 10. Stock solutions at $0.04$~g cm$^{-3}$ were prepared and stored in the dark at 4$^\circ$C for further use. The acoustic contrast for USV measurements was provided by monodisperse micrometric polystyrene latex particles (Microparticles GmbH, diameter 9 $\mu$m) added to the copolymer solution at a weight fraction of 0.01 \cite{manneville04}. To avoid settling of the particles, the solution was first seeded at room temperature. After the dispersion of the latices by vigorous stirring, the solution was brought to about 36$^\circ$C to increase its zero-shear viscosity. In these conditions, the latex particles remained suspended in the solution for the whole duration of the experiments, typically a few hours.

\subsection{Rheology}

Rheological measurements were performed with two home-built setups based on commercial rheometers. They have been described elsewhere \cite{waton04,manneville04}. In both cases the rheometers are stress-controlled but have a feed-back loop that allows strain-controlled measurements. In the temperature range 30$^\circ$C--40$^\circ$C, the temperature regulation of the solution is straightforward and the evaporation is minimized thanks to the use of solvent traps.

In the case of the rheometer coupled with small-angle light scattering \cite{waton04}, the Couette device is operated by a RS1 rheometer (Thermo Haake). The inner rotor has an outer radius $R_{1}=26$~mm and the outer stator an inner radius $R_{2}=27$~mm. Both are made out of quartz. Thermostated water circulating in a jacket surrounding the stator regulates the temperature to within $\pm$ 0.05$^\circ$C, which is measured by a Pt resistor located in the bottom aluminum plate of the jacket.

The setup coupled with USV has very similar characteristics \cite{manneville04}. The inner rotor is operated by an AR1000 rheometer (TA Instruments). The rotor ($R_{1}=24$~mm) and the stator ($R_{2}=25$~mm) are made out of Plexiglas. The stator is surrounded by water whose temperature is kept constant to within $\pm$ 0.1$^\circ$C. The temperature is measured in the circulating bath and the offset with the actual temperature in the Couette cell is estimated to be about 0.5--1$^\circ$C in the range of temperatures probed in these experiments ($36<T<41^\circ$C).

Thus the experimental conditions of the shear flow can be well enough reproduced in the two setups to allow for a comparison of the results. However these conditions are not strictly identical. In particular, the different materials of the Couette devices might influence wall slip and small differences in the actual temperature of the samples will play an important role since temperature controls crucially the micellar growth.

\begin{figure}
\scalebox{0.85}{\includegraphics{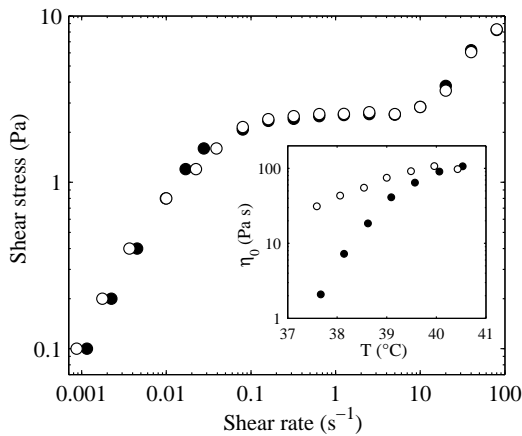}}
\caption{\label{fig:fig1} Influence of latex particles on the rheology of the solutions. Flow curve measured with the setup of Ref.~\cite{waton04} on the P84 solution in brine without particles ($\bullet$) and with particles at 0.01 weight fraction ($\circ$) at a temperature $T=40.2^\circ$C. Inset: zero-shear viscosity $\eta_0$ as a function of the temperature $T$ for the two solutions.}
\end{figure}

We checked for the influence of added latex particles on the rheology with the setup of Ref.~\cite{waton04}. Figure~\ref{fig:fig1} compares the flow curve at a given temperature $T=40.2^\circ$C and the evolution of the zero-shear viscosity with temperature measured in the copolymer solution without latices to the same data obtained in the solution seeded with latices. Although the global temperature evolution is well reproduced in the two samples, the solution with added particles has a much higher zero-shear viscosity for a given temperature in the growth regime (see inset of Fig.~\ref{fig:fig1}). At $T=40.2^\circ$C, close to the maximum of viscosity, the difference in the zero-shear viscosity is less marked. The difference in viscosity even vanishes progressively as the shear rate is increased so that the value and the boundaries of the stress plateau are almost unaffected by the presence of latices. Thus, although the behavior in the non-Newtonian regime, which is of interest in this paper, appears to remain unchanged by the addition of latex particles, the zero-shear viscosity cannot be used as an internal temperature calibration to correct for small differences in actual temperatures. Nevertheless, a qualitative comparison can still be achieved between the previous \cite{waton04} and the present results.

In the following all the results were obtained with the setup of Ref.~\cite{manneville04}. The combined rheological and velocimetry experiments were performed by imposing a constant shear rate to the solutions for increasing shear rate steps. The values reported for the stress correspond to its average value across the gap.

\subsection{Ultrasonic speckle velocimetry}
\label{sec:usv}


The USV setup has been described at length in Ref.~\cite{manneville04}. This technique is based on backscattering from acoustic impedance inhomogeneities suspended in a fluid medium. When the liquid contains a lot of scatterers per unit volume, the backscattered signal results in an ultrasonic speckle built from the interferences of all the backscattered waves. To discriminate between different scatterers in space and to measure velocity profiles in a flow, a solution is to use short acoustic pulses that generate a series of backscattered echoes. The arrival times of the echoes are then directly linked to the positions of the scatterers along the acoustic beam. Cross-correlation of speckle signals corresponding to successive pulses yields the displacement of the scatterers projected along the acoustic axis. By using pulses with central frequencies larger than 20~MHz, velocity profiles can be measured in complex fluids sheared between two plates separated by a gap $e=1$~mm with a spatial resolution of about 40~$\mu$m. The pulse repetition frequency is tuned according to the rotation speed of the rotor so that the maximum displacement between two successive pulses remains smaller than half the acoustic wavelength.

Velocity profiles were recorded after an equilibration time of 30 minutes. For $\dot{\gamma}<15$~s$^{-1}$, an ``individual'' velocity profile corresponds to an average over 1000 successive pulses. For a given number of pulses, the total averaging time is inversely proportional to the applied shear rate \cite{manneville04}. Typically, in the present experiments, for a small shear rate of $\dot{\gamma}\approx$ 0.1~s$^{-1}$, an ``individual'' velocity profile was measured in 250~s. For $\dot{\gamma}\ge 15$~s$^{-1}$, we used averages over 20 bursts of 20 pulses. The period of the burst sequence was chosen so that the total averaging time for an individual velocity profile was equal to 1~s whatever $\dot{\gamma}\ge 15$~s$^{-1}$. The ``time-averaged'' velocity profiles shown below in Figs.~\ref{fig:fig4} and \ref{fig:fig7} are then obtained by averaging individual velocity profiles during about 15~minutes i.e. over only a few individual velocity profiles for the lowest shear rates to several hundreds as soon as $\dot{\gamma}\ge 15$~s$^{-1}$. For the purpose of the discussion below, it is important to keep in mind that USV measures the projection of the velocity vector along the acoustic axis. The reader is referred to Ref.~\cite{manneville04} for full technical details.

\section{Results}
\label{sec:res}

\subsection{$T=37.0^\circ$C : homogeneous flow with wall slip}
\label{sec:t37}

\begin{figure}
\scalebox{0.85}{\includegraphics{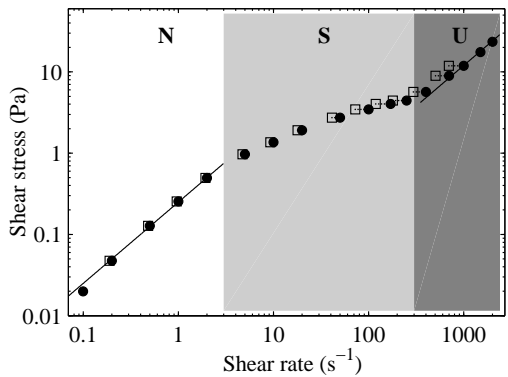}}
\caption{\label{fig:fig3} Flow curve at $T=37.0^\circ$C. The full symbols ($\bullet$) correspond to the engineering data $\sigma(\dot{\gamma})$ while the open symbols ($\square$) represent the ``true'' flow curve $\sigma(\dot{\gamma}_{true})$ obtained once wall slip is taken into account (see text). The solid lines are  $\sigma=0.25\dot{\gamma}$ in regime (N) and $\sigma=0.012\dot{\gamma}$ in regime (U).}
\end{figure}

\begin{figure}
\scalebox{0.85}{\includegraphics{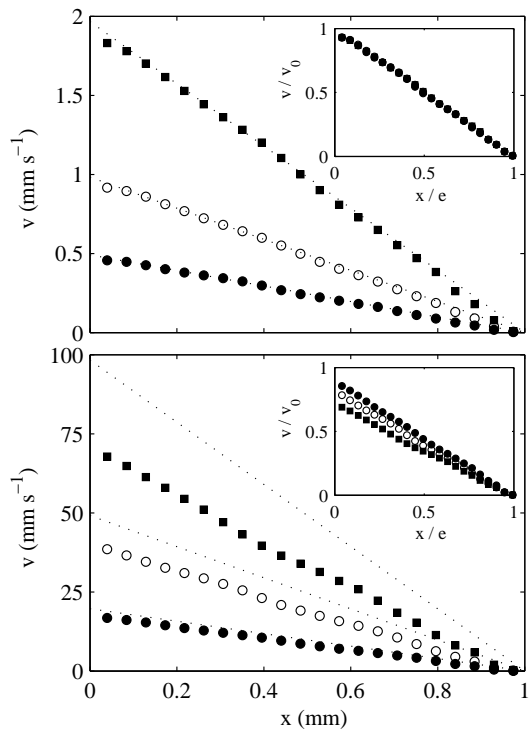}}
\caption{\label{fig:fig4} Time-averaged velocity profiles at $T=37.0^\circ$C. The standard deviation is of the order of the marker size. The dotted lines are the velocity profiles expected in a Newtonian fluid without wall slip. Inset: dimensionless data $v_{s}/v_{0}$ vs $x/e$. Top :  $\dot{\gamma}$ = 0.5 ($\bullet$), 1 ($\circ$), and 2~s$^{-1}$ ($\blacksquare$). Bottom : $\dot{\gamma}$ = 20 ($\bullet$), 50 ($\circ$), and 100~s$^{-1}$ ($\blacksquare$).}
\end{figure}

Figure \ref{fig:fig3} shows the flow curve measured by the rheometer at $T=37.0^\circ$C (hereafter referred to as the ``engineering'' flow curve). As reported earlier \cite{waton04}, it shows an inflexion point but no stress plateau. At low shear rates, a zero-shear viscosity of 0.25 Pa~s is measured. Figure~\ref{fig:fig4} shows the corresponding  time-averaged velocity profiles $v(x)$ measured for different applied shear rates, where $x$ is the distance from the rotor. The stator is located at $x=e=R_2-R_1=1$~mm. For the whole range of shear rates investigated here, the time-averaged velocity profiles correspond to those expected for a Newtonian fluid and the time-averaged velocity gradient is constant across the whole gap.

\begin{figure}
\scalebox{0.85}{\includegraphics{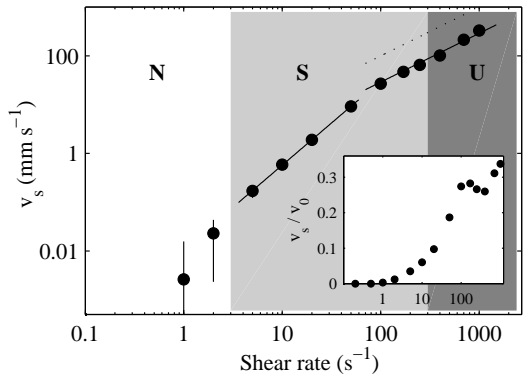}}
\caption{\label{fig:fig5} Time-averaged slip velocity $v_s$ at the rotor at $T=37.0^\circ$C vs applied shear rate $\dot\gamma$. The error bars correspond to the standard deviation of the measurements. The solid lines have slopes 1.7 and 1 (see discussion in Sect.~\ref{sec:wallslip}). The dashed line is $v_s=v_0$. For $\dot{\gamma}<1$~s$^{-1}$, slip velocities are too small to show in logarithmic scales. Inset: relative slip $v_s/v_0$ at the rotor vs applied shear rate $\dot\gamma$.}
\end{figure}

However, for applied shear rates higher than about 3~s$^{-1}$, there is a noticeable difference between the velocity $v_0=\dot\gamma e$ imposed by the rotor and the true velocity of the fluid at the rotor ($x=0$), i.e., wall slip occurs at the rotor. A quantitative estimate of the slip velocity is obtained through a linear fit $\tilde{v}(x)$ of the experimental data close to the inner wall over $x=0.08$--0.20~mm. The slip velocity $v_{s}$ at the rotor is then simply given by $v_s=v_0-\tilde{v}(0)$. As seen in Fig.~\ref{fig:fig5}, wall slip does not set in abruptly for a well defined shear rate but the slip velocity rather appears to increase smoothly over the whole range of applied shear rates. Note the error bars on the first data points which indicate an experimental uncertainty on the determination of $v_s$ of about 0.01~mm s$^{-1}$. As long as $\dot\gamma<3$~s$^{-1}$, wall slip thus remains hardly detectable and is buried in experimental noise. The relative slip $v_s/v_0$ increases until it reaches about 30\% for $\dot\gamma>100$~s$^{-1}$ (see inset of Fig.~\ref{fig:fig5}). For the purpose of further discussion we notice that no wall slip is observed at the stator ($x=1$).

Finally, as long as the imposed shear rate remains smaller than 300~s$^{-1}$, the flow is stationary as shown by the very small standard deviation of the velocity measurements (of the order of the marker size in Fig.~\ref{fig:fig4}). For shear rates higher than 300~s$^{-1}$, the flow reaches a new, unstable regime, whose signature is the considerably larger statistical fluctuations of the instantaneous velocity profiles around their average value (velocity profiles not shown, see Sect.~\ref{sec:u} for more details on this unstable flow regime). The flow profile becomes non-stationary although its mean shape for long time averages remains close to Newtonian in the bulk with about 30\% wall slip at the rotor.

In Fig.~\ref{fig:fig3} we marked by different background grey values the regions corresponding to Newtonian behavior (N), noticeable wall slip (S), and unstable flow (U). Figure~\ref{fig:fig3} also shows the ``true'' flow curve, i.e., the shear stress $\sigma$ plotted against the shear rate corrected for wall slip $\dot\gamma_{true}=\dot\gamma-v_s/e$ hereafter called the ``true'' shear rate (open symbols). Here and in the following the limits of the different flow regimes are defined according to the true shear rates.

\subsection{$T=39.4^\circ$C : complex banded flow}
\label{sec:t39}

We now turn to measurements performed at a higher temperature where the flow curve was previously shown to present a stress plateau \cite{waton04}. We first give a general overview of the rheological and flow behaviors and then describe more precisely the various flow regimes observed at $T=39.4^\circ$C.

\subsubsection{Rheological and flow behavior overview}

\begin{figure}
\scalebox{0.85}{\includegraphics{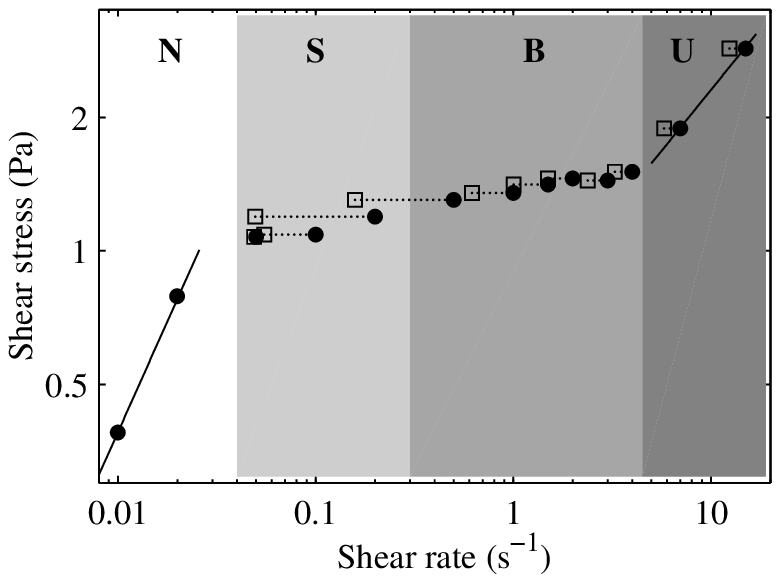}}
\caption{\label{fig:fig6}Flow curve at $T=39.4^\circ$C. The full symbols ($\bullet$) correspond to the engineering data $\sigma(\dot\gamma)$ while the open symbols ($\square$) represent the ``true'' flow curve $\sigma(\dot\gamma_{true})$ obtained once wall slip is taken into account. The solid lines are $\sigma=39\dot\gamma$ in regime N and $\sigma=0.65\dot\gamma^{0.55}$ in regime U.}
\end{figure}

\begin{figure}
\scalebox{0.85}{\includegraphics{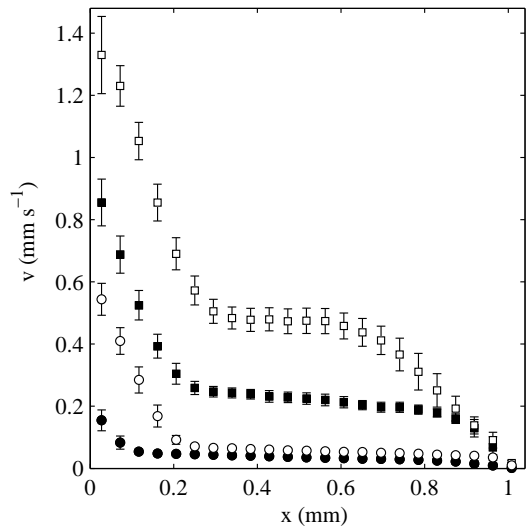}}
\caption{\label{fig:fig7} Time-averaged velocity profiles at $T=39.4^\circ$C in the shear banding regime noted (B) in Fig. \ref{fig:fig6}. $\dot{\gamma}$ = 0.5 ($\bullet$), 1 ($\circ$), 1.5 ($\blacksquare$), and 2~s$^{-1}$ ($\square$). The error bars correspond to the standard deviation of the measurements. Note that the corresponding slip velocities can be easily estimated from the applied shear rate and from the fluid velocity measured at the rotor using $v_s=\dot\gamma e-v(x=0)$.}
\end{figure}

As shown in Fig.~\ref{fig:fig6}, the flow curve at $T=39.4^\circ$C exhibits an almost flat stress plateau at $\sigma\approx 1.2$~Pa over more than two decades separating a Newtonian branch (N) for $\dot{\gamma}\lesssim 0.03$~s$^{-1}$ ($\eta_{0}$ $\approx$ 39 Pa~s) and a second increasing branch for $\dot{\gamma}\gtrsim 5$~s$^{-1}$ corresponding to unstable flow (U). From velocity profile measurements, the stress plateau can be divided into two parts: a first regime (S) for $0.03<\dot\gamma<0.3$~s$^{-1}$ where only wall slip is observed, and a new regime (B) for $0.3<\dot\gamma<5$~s$^{-1}$ where time-averaged velocity profiles reveal a three-band structure together with wall slip (see Fig.~\ref{fig:fig7}). Velocity profiles in regimes (N), (S), and (U) (not shown) are widely similar to those measured in the corresponding regimes at $T=37.0^\circ$C. In particular, the high shear unstable branch (U) is again characterized by very large temporal fluctuations ($\delta v/v\gtrsim 50\%$) around an average homogeneous velocity profile with significant wall slip at the rotor. The rheological and flow behaviors were found to be qualitatively the same for all investigated temperatures $T\geq 38^\circ$C.

\subsubsection{Rheological signals}

\begin{figure}
\scalebox{0.85}{\includegraphics{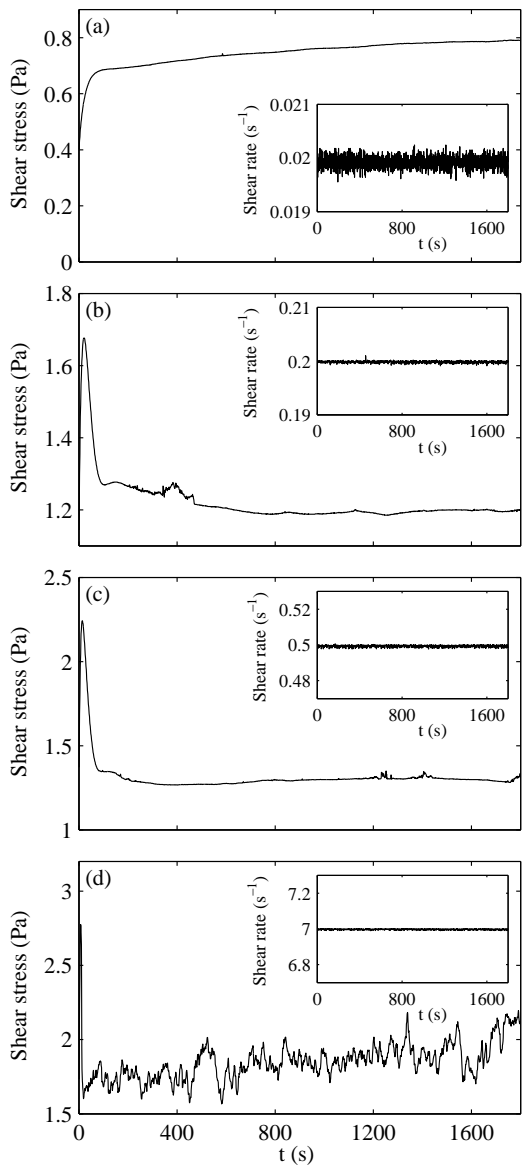}}
\caption{\label{fig:fig8} Rheological time series $\sigma(t)$ recorded at $T=39.4^\circ$C for four different imposed shear rates. (a) Newtonian regime (N): $\dot\gamma=0.02$~s$^{-1}$. (b) Wall slip regime (S): $\dot\gamma=0.2$~s$^{-1}$. (c) Shear banding regime (B): $\dot\gamma=0.5$~s$^{-1}$. (d) Unstable regime (U): $\dot\gamma=7$~s$^{-1}$. The insets show that the feedback loop of the rheometer keeps the shear rate $\dot\gamma(t)$ constant to within 2\% for $\dot\gamma=0.02$~s$^{-1}$, 0.6\% for $\dot\gamma=0.2$~s$^{-1}$, and  0.3\% for $\dot\gamma=0.5$ and 7~s$^{-1}$.
}
\end{figure}

Figure~\ref{fig:fig8} shows that in regimes (N), (S), and (B), steady state is reached within about 10 minutes. A stress overshoot in the temporal evolution of the shear stress $\sigma(t)$ marks the onset of nonlinear behavior for regimes (S), (B), and (U). In regime (U), the instability of the flow is revealed by the significant increase of the temporal fluctuations of $\sigma(t)$: typically $\delta\sigma/\sigma\approx 20\%$ whereas $\delta\dot{\gamma}/\dot{\gamma}\leq 0.3\%$. The Fourier spectrum of $\sigma(t)$ exhibits a single characteristic frequency  corresponding to that of the rotor rotation in an otherwise $1/f$ behavior. The evolution of the relative fluctuations of the applied shear rate depicted in the insets of Fig.~\ref{fig:fig8} reflects the increasing difficulty to control the shear rate as it becomes very low. We could not detect any significant temporal correlation between the fluctuations of the applied shear rate and those of the measured stress.

\subsubsection{Slip velocity}
\label{sec:s}

\begin{figure}
\scalebox{0.85}{\includegraphics{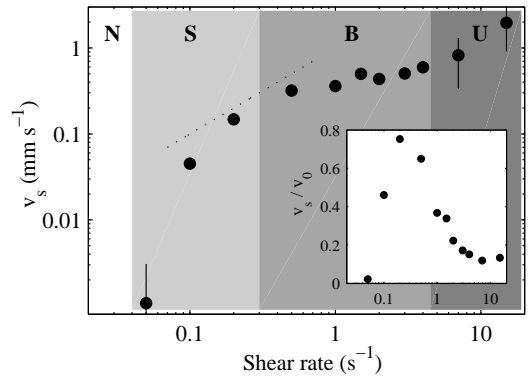}}
\caption{\label{fig:fig9}Time-averaged slip velocity $v_s$ at the rotor at $T=39.4^\circ$C vs applied shear rate $\dot\gamma$. The error bars correspond to the standard deviation of the measurements. The dashed line is $v_s=v_0$.  Inset: relative slip $v_s/v_0$ at the rotor vs applied shear rate $\dot\gamma$.}
\end{figure}

Once again no slip is detected at the stator for $T=39.4^\circ$C. The evolution of the slip velocity at the rotor with the imposed shear rate is shown in Fig.~\ref{fig:fig9}. It strikingly differs from that observed at $T=37.0^\circ$C (Fig.~\ref{fig:fig5}). After a sharp increase in regime (S), $v_s$ almost saturates in the shear banding regime (B). This behavior corresponds to a decrease of the relative slip $v_{s}/v_{0}$ as $\dot\gamma$ is increased (see inset of Fig.~\ref{fig:fig9}) which starts at the onset of shear banding when $v_{s}/v_{0}\approx 0.8$. The same criterion was fulfilled at all temperatures where shear banding could be observed ($T\geq 38^\circ$C). Note that for $T=37.0^\circ$C the ratio $v_{s}/v_{0}$ remains always smaller than 0.3 although the absolute value of the slip velocity can be orders of magnitude larger at the lower temperature (see Fig.~\ref{fig:fig5}).

\subsubsection{Shear banding regime}
\label{sec:b}

\begin{figure}
\scalebox{0.85}{\includegraphics{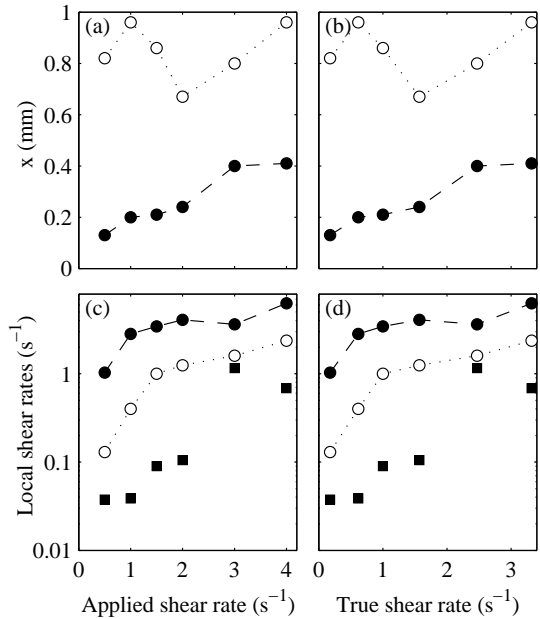}}
\caption{\label{fig:fig10} Analysis of the three-band velocity profiles recorded in regime (B) at $T=39.4^\circ$C. Positions of the interfaces inside the gap as a function (a) of the applied shear rate and (b) of the true shear rate. Local shear rates in the highly sheared bands at the rotor ($\bullet$) and at the stator ($\circ$) and in the central weakly sheared band ($\blacksquare$) as a function (c) of the applied shear rate and (d) of the true shear rate.}
\end{figure}

The most striking feature of Fig.~\ref{fig:fig7} is the presence of three shear bands in the gap. Figure~\ref{fig:fig10} shows the location of the two interfaces across the gap and the local shear rate values within the bands as a function of the true shear rate. Such data are easily extracted from piece-wise linear fits of the time-averaged velocity profiles in regime (B). The width of the high shear band close to the rotor is seen to increase with the shear rate whereas no systematic trend can be given for the band at the rotor. Plotting the same data as a function of the local shear stresses at the interfaces does not provide evidence for any ``stress selection'' rule either. Moreover the central band remains very weakly sheared for most of the imposed shear rate values. The shear rate inside the band at the rotor is always at least twice as large as that in the band at the stator. These results clearly point to a complex shear banding scenario which will be discussed below in Sect.~\ref{sec:sb}. Note that temporal fluctuations of the velocity field remain small in regime (B) but not quite negliglible: standard deviations in Fig.~\ref{fig:fig7} are $\delta v/v\approx 5$--10\% typically. A detailed study of these fluctuations is left for future work.

\subsubsection{Unstable flow regime}
\label{sec:u}

\begin{figure}
\scalebox{0.85}{\includegraphics{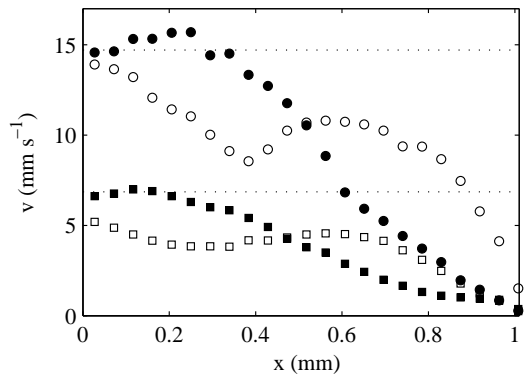}}
\caption{\label{fig:fig11} Individual velocity profiles recorded at $T=39.4^\circ$C for 
$\dot{\gamma}=7$~s$^{-1}$ ($\blacksquare$,$\square$) and $\dot{\gamma}=15$~s$^{-1}$ ($\bullet$,$\circ$). The dotted lines indicate the corresponding rotor velocities $v_{0}$.}
\end{figure}

\begin{figure}
\scalebox{0.85}{\includegraphics{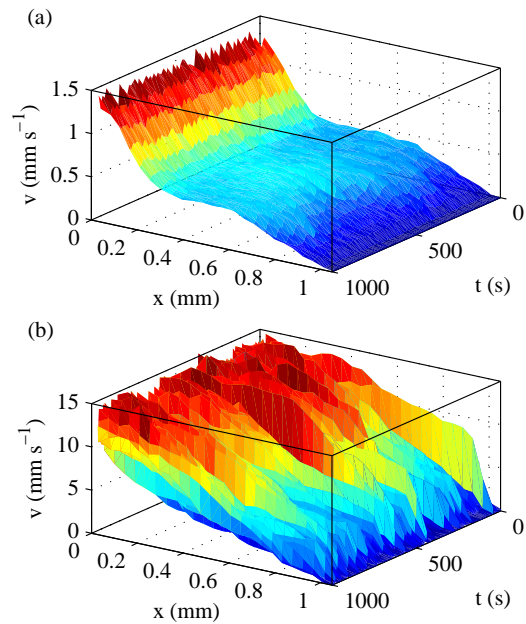}}
\caption{\label{fig:fig12} Spatio-temporal representation of the velocity profiles $v(x,t)$ at $T=39.4^\circ$C for (a) $\dot{\gamma}=2$~s$^{-1}$ and (b) $\dot{\gamma}=15$~s$^{-1}$ (color online).}
\end{figure}

Examples of instantaneous velocity profiles measured within about 1~s in regime (U) are plotted in Fig.~\ref{fig:fig11}. They clearly indicate that the flow is strongly non-stationary. The spatio-temporal representations of the velocity profiles $v(x,t)$ shown in Fig.~\ref{fig:fig12} illustrate the drastic difference between the stationary flow observed in regime (B) and the unstable flow regime (U) where huge temporal fluctuations of $v(x)$ are recorded. Finally we shall argue in Sect.~\ref{sec:unstable} that the flow becomes cannot remain purely tangential in regime (U).

\section{Discussion}
\label{sec:discuss}

\subsection{Wall slip}
\label{sec:wallslip}


In complex fluids like colloidal suspensions, the possible break-down of the no-slip boundary condition has been shown for a long time \cite{mooney31,barnes95}. Depending on the net interaction between the walls and the suspended colloids, different mechanisms are involved. For neutral or repulsive walls, wall depletion occurs and creates a lubricating layer of the suspending fluid close to the wall. The equilibrium concentration profile at rest can be further modified by the flow, in particular when the particles are soft and deformable \cite{leal80}. This also occurs in polymer solutions \cite{agarwal94}. For attractive walls, partly covered with adsorbed colloids, wall slip could be expected to be rather uncommon. One notable exception is the case of entangled polymer solutions and melts when entanglements between an adsorbed layer and the bulk system are present. Such phenomena have been investigated both theoretically and experimentally \cite{brochard92,brochard96,migler92,durliat97,leger97,hervet03}.

Near field laser velocimetry experiments on poly(dimethylsiloxane) (PDMS) melts sheared between sliding plates grafted with PDMS chains have shown three different slip regimes \cite{migler92,durliat97,leger97,hervet03}. At low and high plate velocities, the slip velocity $v_{s}$ is proportional to the wall velocity $v_{0}$, with  a quasi plug flow, $v_{s}\approx v_{0}$, at high velocities. In between these two limits, a nonlinear friction regime is observed, where $v_{s}$ increases more rapidly than linearly with $v_{0}$. In this intermediate regime, the slip length $b=e v_s/(v_0-v_s)$, i.e., the distance at which the velocity profile extrapolates to the wall velocity, increases almost linearly with the slip velocity ($b\propto v_s^\alpha$ with $\alpha=0.8$--1.2). These results have been explained in terms of a dynamic decoupling between the surface and the bulk polymer chains \cite{brochard92,brochard96}. More precisely, in the case of grafted chains, such a decoupling arises from the competition between the strongly entanglement-dependent frictional force that stretches the chains and the entropic elastic force that recoils them. The resulting steady-state extension of the surface and bulk chains increases with the shear rate until the stretched chains are close to disentangle. This is the onset of the intermediate slip regime which lasts until the shear rate becomes larger than the characteristic frequency at which the chains can reform new entanglements. Then the surface and the bulk chains become fully disentangled, the friction decreases by two orders of magnitude, and the slip velocity is high so that $v_{s}\approx v_{0}$.

\begin{figure}
\scalebox{0.85}{\includegraphics{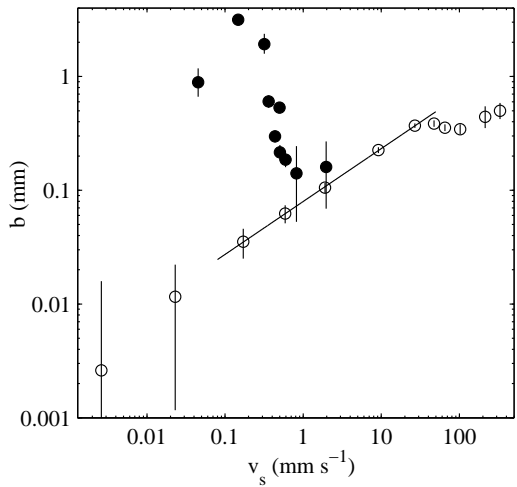}}
\caption{\label{fig:fig13} Slip length $b$ as a function of slip velocity $v_s$ at the rotor for $T=37.0^\circ$C ($\circ$) and $T=39.4^\circ$C ($\bullet$). The solid line is a power law with exponent 0.47.}
\end{figure}

At the lowest temperature investigated in the present work, $T=37.0^\circ$C, the behavior of the slip velocity as a function of shear rate is strongly reminiscent of the one reported for PDMS melts in contact with PDMS grafted surfaces \cite{leger97}. Indeed, as shown by the lines in Fig.~\ref{fig:fig5}, a weak slip regime, with a behavior obscured by experimental noise, and a strong slip regime, where $v_s\propto\dot\gamma$, are separated by a region where the slip velocity increases faster than linearly with the imposed shear rate ($v_s\propto\dot\gamma^{1.7}$). In the strong slip regime, however, the slip velocity never reaches the limiting value $v_{s}=v_{0}=\dot\gamma e$ (indicated by a dashed line in Fig.~\ref{fig:fig5}) but rather saturates at $v_s\approx 0.3 v_0$. The corresponding slip length is shown in Fig.~\ref{fig:fig13}. Due to the large uncertainty at small shear rates, we cannot give a reliable value for the slip length $b$ in the weak slip regime, which cannot be distinguished from perfect Newtonian behavior using the present measurements. Still, although the power law behavior in our data is significantly different ($b\sim v_s^\alpha$ with $\alpha\simeq 0.5$), the slip lengths of several hundred of microns observed in the intermediate regime are close to those reported on PDMS (compare, e.g., Fig.~7 of Ref.~\cite{leger97} to Fig.~\ref{fig:fig5} of the present paper).

The above similarity between our measurements at $T=37.0^\circ$C and previous results on PDMS melts sheared between PDMS grafted plates prompts us to interpret wall slip in our experiments as a result of the dynamic decoupling scenario described above. Of course, in our case, micelles are not grafted at the cell walls. Nevertheless we may assume that PEO--PPO--PEO wormlike micelles adsorb on the surfaces. Indeed at least one experimental study has clearly demonstrated the adsorbtion of wormlike micelles on a substrate for poly(styrene)-poly(isoprene) diblock copolymers in heptane, a selective solvent for poly(isoprene) \cite{larue04}. No similar results have been reported for wormlike micelles made of PEO--PPO block copolymers. However it has been shown that spherical micelles formed from diblock PEO--PPO copolymers close to the bulk cmc adsorb on hydrophilic or hydrophobic silica substrates while retaining their shape \cite{connel03,hamley04}. Thus in principle we can expect similar effects in our system, namely adsorbtion of wormlike micelles on the walls. 

For adsorbed polymers, the possibility of tearing off of the surface chains due to the shear stress should also play a role and could explain the difference in the scaling exponents and in the saturation value of $v_s$. To the best of our knowledge, no theoretical prediction is available in the case of adsorbed chains. Moreover, in the case of adsorbed surface wormlike micelles, the scission of the micelles should also provide a mechanism for disentanglement. If the breaking time $\tau_{b}$ is much smaller than the reptation time $\tau_{R}$, one expects to recover in a first approximation $\tau\sim (\tau_{b}\tau_{R})^{1/2}$ for the characteristic relaxation time \cite{cates87}. Thus the transition to the strong slip regime is expected to occur for $\dot{\gamma}\tau\gtrsim 1$. In our case, however, the bulk relaxation time $\tau$ at $T=37.0^\circ$C is about 1~s \cite{waton04} so that the onset of strong slip corresponds to $\dot{\gamma}\tau\approx 100$. Such a discrepancy is easily explained by the fact that adsorbed micelles should have a much smaller contour length left available to entangle with the bulk micelles. A more thorough investigation of the proposed scenario for wall slip at $T=37.0^\circ$C is beyond the scope of this paper.

On the other hand, the $v_s$ vs $\dot\gamma$ data for the higher temperature $T=39.4^\circ$C display a plateau-like behavior in the intermediate regime between weak and strong slip regimes (see Fig.~\ref{fig:fig9}). The difference with the lower temperature is even more striking when the slip length is considered (see Fig.~\ref{fig:fig13}). At $T=39.4^\circ$C, after a large initial increase, $b$ suddenly drops as the system enters the shear banding regime. Such a decrease was also observed in a lyotropic lamellar phase in the shear banding regime \cite{salmon03b}. This behavior was interpreted in terms of a decrease of the thickness $h$ of the lubricating layers that were assumed to be composed of perfectly aligned membranes. Indeed if one notes $\eta_s$ the viscosity of the fluid inside the slip layer and if one assumes that $h\ll e$ then stress continuity imposes $\eta_s v_s/h=\eta (v_0-v_s)/e$, where $\eta$ is the bulk viscosity. In other words, one has $h=b\eta_s/\eta$. Assuming $\eta_s$ to remain constant and equal to the solvent (brine) viscosity, a decrease of $h$ in the shear banding regime was attributed to solvent migration from the walls to the shear-induced low viscosity fluid close to the rotor \cite{salmon03b}. In the present case of PEO--PPO--PEO wormlike micelles, a constant $\eta_s$ seems like a rather strong assumption since one expects progressive disentanglement in the slip layer unless the strong slip regime is reached. An alternative explanation could be that the rheology of the slip layer itself presents a non-monotonic behavior leading to instability. Such a behavior was recently observed and linked to a stick-slip instability in nanotribology experiments on confined surfactant layers adsorbed on the mica surfaces of a surface force apparatus \cite{drummond03,drummond04}. Finally we shall see below that the slip regime (S) in Figs.~\ref{fig:fig6} and \ref{fig:fig9} is also compatible with a very thin high shear band unresolved by our velocimetry setup, so that a definite interpretation of wall slip at $T=39.4^\circ$C is still an open question.

To summarize our discussion on wall slip, the measurements performed at $T=37.0^\circ$C and presented in Sect.~\ref{sec:t37} are broadly consistent with the behavior expected for the dynamic decoupling at the interface between bulk and adsorbed surface chains. This supports the hypothesis that adsorbed wormlike micelles could be involved in the observed slip behavior. At $T=39.4^\circ$C, the slip phenomena are markedly different, in correlation with the observation of shear banding in the bulk at this temperature, although the interplay between slip and shear banding remains unclear.

\subsection{Shear banding}
\label{sec:sb}

Shear-induced separation of complex fluids flow into bands of different viscosities has been predicted and interpreted for almost two decades in terms of an underlying non-monotonic flow curve which leads to the presence of a stress plateau at $\sigma^*$ between two critical shear rates $\dot\gamma_1$ and $\dot\gamma_2$ in the rheological experiments \cite{mcleish86,spenley93,rehage91}. In the most simple scenario for shear banding, two or more shear bands of viscosities $\eta_1=\sigma^*/\dot\gamma_1$ and $\eta_2=\sigma^*/\dot\gamma_2$ coexist along the stress plateau and the proportion $\alpha$ of the highly sheared fluid is given by the so-called ``lever rule'' $\alpha=(\dot\gamma-\dot\gamma_1)/(\dot\gamma_2-\dot\gamma_1)$: as the imposed shear rate $\dot\gamma$ is increased from $\dot\gamma_1$, the high shear bands progressively invade the whole sample until $\dot\gamma=\dot\gamma_2$. The plateau value was predicted to be $\sigma^*=0.67 G_0$ (with $G_0$ the plateau modulus) \cite{spenley93}, which was indeed observed in some experiments on wormlike micelles \cite{rehage91,khatory93,berret94}. In the concentric cylinder geometry, the (small) stress inhomogeneity results in the presence of only two bands with the high shear band close to the inner cylinder. Moreover the lever rule was shown to hold in CPCl-NaSal semidilute wormlike micellar solutions \cite{salmon03,lerouge04,hu05}.

More refined models involved constitutive equations based on the frame-invariant Gordon-Schowalter derivative \cite{larson88}. In particular, the inclusion of diffusive terms in such models were shown to be a necessary ingredient to obtain stress selection in planar geometries \cite{yuan99,radu00,olmsted00,lu00}. The possibility of non-uniform concentration in the gap was introduced by the means of a two-fluid description and the coupling between flow and concentration \cite{yuan02,fielding03,fielding03b,fielding03c}. This last model should be well adapted to describe our experimental system since it incorporates the flow-concentration coupling that can explain butterfly isointensity patterns parallel to the flow direction, measured earlier at $T\lesssim37^\circ$C \cite{waton04}, and allows for a phase separation in the system.

The flow curve of Fig.~\ref{fig:fig6} looks typical of a simple shear banding phenomenon with $\sigma^*\approx 1.2$~Pa, $\dot{\gamma}_1\approx 0.03$~s$^{-1}$, and $\dot{\gamma}_2\approx 5$~s$^{-1}$. With a plateau modulus $G_0\approx 1.8$~Pa \cite{waton04}, the observed stress plateau is even nicely predicted by the most simple toy model, $\sigma^*=0.67 G_0$ \cite{spenley93}. Note also that apparent wall slip in regime (S) may also be interpreted in the framework of a simple shear banding scenario. Indeed, if the lever rule is valid, the width of the high shear band remains smaller than the USV spatial resolution $\delta_0\approx 40~\mu$m as long as $\dot\gamma<\dot\gamma_1+\delta_0(\dot\gamma_2-\dot\gamma_1)/e\approx 0.23$~s$^{-1}$. Thus due to the huge difference between $\eta_1\approx 40$~Pa~s and $\eta_2\approx 0.2$~Pa~s, the high shear band may be too thin to be detected over a significant fraction of the stress plateau, which would be consistent with the slip regime observed for $\dot{\gamma}=0.03$--0.3~s$^{-1}$.

However the velocity profiles measured along the stress plateau at $T=39.4^\circ$C and shown in Fig.~\ref{fig:fig7} clearly point to a much more complex shear banding scenario than predicted by even the most sophisticated models  \cite{yuan02,fielding03,fielding03b,fielding03c} for the following reasons. (i)~Stationary states with three shear bands are recorded over the whole stress plateau. (ii)~Strong wall slip is detected at the rotor. (iii)~The various shear rates inside the bands do not remain constant as the applied shear rate is increased. (iv)~Even if the true shear rate is considered, the lever rule does not hold. To the best of our knowledge, even if transient three-banded velocity profiles were already reported in wormlike micelles \cite{becu04b}, this is the first evidence for such profiles in the steady state. 
This observation should undoubtedly be confronted to the very recent model of Ref.~\cite{rossi06}. Based on a bead-spring mechanism with non-affine motion \cite{cook04}, this model incorporates the same basic ingredients as the two-fluid diffusive Johnson-Segalman model \cite{yuan02,fielding03,fielding03b,fielding03c} and was shown to predict various flow structures depending on the boundary conditions. If the conformation flux across the walls is set to zero (Neumann condition), two-banded steady states are found whereas when the micelles are aligned at the walls (Dirichlet condition) the model predicts a stationary three-band structures very similar to that observed in our experiments. Moreover, in the latter case, strong depletion is observed at both walls. Such a model may provide some keys to the unusual shear banding reported in the present study and more specifically to points (i) and (ii) above. Note that a similar approach was proposed in the framework of yield stress fluids and that similar results with multi-banded states depending on boundary conditions were obtained in the fluidity model of Ref.~\cite{picard02}. Finally, concerning issues (iii) and (iv), let us note that a complex shear banding scenario with multiple successive transitions where the local shear rates do not remain constant and the lever rule does not hold was also reported very recently for solutions of self-assembled polymers \cite{vdg06}. Contrary to what we observed in regime U (see below), we did not notice any evidence for two- or three-dimensional flows in regime B.

\subsection{Unstable flows}
\label{sec:unstable}

\begin{figure}
\scalebox{0.85}{\includegraphics{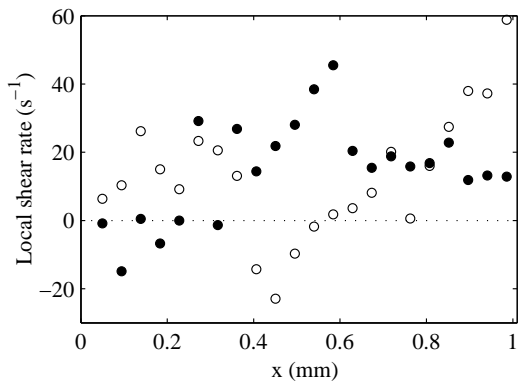}}
\caption{\label{fig:fig14} Local shear rate $\dot\gamma(x)$ computed from the individual velocity profiles recorded at $T=39.4^\circ$C for $\dot{\gamma}=15$~s$^{-1}$ and shown in Fig.~\ref{fig:fig11}.}
\end{figure}

In all cases investigated in this work, the flow strongly fluctuates in time as soon as the high shear branch of the flow curve is reached. The individual velocity profiles of Fig.~\ref{fig:fig11} even show portions with positive slope. More precisely, in spite of scatter due to differentiation, the local shear rate shown in Fig.~\ref{fig:fig14} and computed from
\begin{equation}
\dot\gamma(x)=\frac{v}{r}-\frac{\partial v}{\partial r}\,,
\label{gammaloc}
\end{equation}
where $r=R_1+x$, becomes clearly negative for $x\approx 0.1$--0.2~mm in one case and $x\approx 0.4$--0.5~mm in the other. Since a negative local shear rate is not compatible with material stability, we conclude that the flow cannot remain purely tangential.

Indeed, as recalled in Sect.~\ref{sec:usv}, USV only provides a measurement of the velocity vector projected along the acoustic axis. So far our interpretation of the velocity profiles has relied on the assumption that the flow is purely tangential. Here the only way to account for a negative apparent shear rate is to include a non-zero radial component $v_r$ in the velocity field so that the experimentally measured velocity actually reads 
\begin{equation}
v(x)=v_\theta(x)+\frac{v_r(x)}{\tan\theta}\,,
\label{e.modvx}
\end{equation}
where $v_\theta$ is the tangential component of the velocity and $\theta$ the angle between the acoustic axis and the radial direction. Since $\theta\approx 15^\circ$ \cite{manneville04}, the effect of radial velocity is amplified by a factor of about 4 and, depending on $v_r(x)$, Eq.~(\ref{e.modvx}) may lead to ``negative local shear rates'' when used in Eq.~(\ref{gammaloc}). Moreover the detection of velocities larger than the wall velocity $v_0$ close to the rotor is another strong indication for a non-zero radial component of the velocity (see Fig.~\ref{fig:fig11}). Note that, for $\dot{\gamma}=7$ and 15~s$^{-1}$, the rotation periods of the rotor are about 21~s and 10~s respectively, so that individual velocity profiles (recorded in about 1~s) only correspond to a fraction of the rotor revolution. Thus the present experiments do not allow us to discriminate between a two-dimensional non-axisymmetric flow, i.e. where $v_\theta$ and $v_r$ are functions of both $r$ and $\theta$ but the vertical velocity $v_z$ remains zero, and a fully three-dimensional flow with all three velocity components being non-zero.

As a possible cause for instability, we may first rule out any inertial effect such as the Taylor-Couette instability. Indeed, even on the high shear branch of the flow curve at $T=39.4^\circ$C where $\eta\approx 0.2$~Pa~s and $\dot\gamma\approx 10$~s$^{-1}$, the Reynolds number is $Re=\rho\dot\gamma e^2/\eta\approx 0.05$ and the Taylor number is $Ta=(e/R_1) Re^2\approx 10^{-4}$ so that the flow can be considered as inertialess. In the Newtonian-like case observed at $T=37.0^\circ$C, one has $\eta\approx 10$~mPa~s for the highest effective shear rate $\dot\gamma\approx 10^3$~s$^{-1}$. Although these are only rough estimates, one gets $Ta\approx 400$ which remains smaller than the critical Taylor number 1712 for a Newtonian fluid.

We rather propose to interpret the observed unstable flows as a result of an elastic instability \cite{muller89,larson90,shaqfeh96}. Such an instability is driven by negative normal stress differences that may cause a streamline to become unstable with respect to radial perturbations. As shown by theoretical analyses for viscoelastic fluids, a purely elastic instability may develop when $(e/R_2)^{1/2}Wi >M_c$, where $Wi=\dot\gamma\tau$ is the Weissenberg number and $M_c$ some critical number of order 1 whose exact value depends on the constitutive model assumed for the fluid. For instance, one finds $M_c\approx 5.9$ for both the Upper-Convected Maxwell model \cite{larson90} and the Oldroyd-B model when the viscosity of the solvent is half that of the polymeric part \cite{oztekin96}.

Previous rheological measurements on the PEO--PPO--PEO Pluronic P84 in brine have shown that the relaxation time increases from $\tau\approx 1$~s at $T=37.0^\circ$C to $\tau\approx 25$~s at $T=40.0^\circ$C \cite{waton04}. Assuming $M_c\approx 5.9$, these measurements lead to critical shear rates $\dot\gamma_c\approx 30$~s$^{-1}$ at $T=37.0^\circ$C and $\dot\gamma_c\approx 1.5$~s$^{-1}$ at $T=39.4^\circ$C for the onset of the elastic instability. The velocimetry experiments reveal non-stationary unstable flows above 300~s$^{-1}$ and 5~s$^{-1}$ for $T=37.0^\circ$C and $T=39.4^\circ$C respectively, well above the predicted thresholds for the elastic instability. Our observations are thus consistent with an elastic instability, although the early stages of such an instability, i.e., an initially toroidal stationary flow with Taylor-like vortices that undergo further bifurcation towards periodic states as the rotation speed is increased \cite{muller89,shaqfeh96}, should in principle be observed for $\dot\gamma=30$--300~s$^{-1}$ and $\dot\gamma=1.5$--5~s$^{-1}$ respectively. At $T=37.0^\circ$C Newtonian-like velocity profiles are recorded all along the flow curve and no stationary toroidal flow was observed. At $T=39.4^\circ$C, due to the complexity of the shear-banded velocity profiles, we cannot rule out the presence of a small time-independent toroidal flow along the stress plateau. Moreover the exact viscoelastic properties of the fluid at high shear rates are unknown. In particular, the relaxation time of the sheared fluid is probably much smaller than that measured in the linear regime so that the instability threshold may be pushed towards higher shear rates. Finally the exact value of $M_c$ may differ significantly from 5.9 so that the present discussion can only remain rather qualitative. 

\section{Conclusions}

We have investigated the flow of a PEO--PPO--PEO Pluronic P84 solution in brine using simultaneous velocity profile and rheological measurements. This study completes the small-angle light scattering measurements performed earlier on the same system \cite{waton04}. It confirms the presence of shear-banding and instability in the flow of this triblock copolymer micellar solution. When the rheological flow curve is an increasing function of the shear rate, i.e., for $T<T_c\simeq 37.7^\circ$C, the velocity profiles remain Newtonian but a transition from weak to strong wall slip is detected at the rotor. The evolution of the slip velocity was compared to previous results on PDMS and favors an interpretation in terms of micelles adsorbed at the wall that progressively untangle from the bulk fluid. For $T>T_c$ the flow curve displays a stress plateau that was correlated to the presence of shear banding. However our data clearly reveal a complex shear banding scenario with wall slip, stationary three-band structure and varying local shear rates in the bands. The presence of three shear bands in the gap was discussed in light of recent theoretical predictions accounting for the influence of boundary conditions on bulk behavior. In connection with wall slip, the possibility that micelles get strongly aligned at the walls could be at the origin of a three-band structure. Finally, in all cases, the flow was shown to become strongly non-stationary and either non-axisymmetric or even possibly three-dimensional at high shear rates. Such unstable flows are consistent with the occurence of an elastic instability on the high shear branch of the flow curve although information on the structure and rheology of the fluid in this regime is missing.

\begin{acknowledgments}
The authors wish to thank L. B\'{e}cu for technical help on the experiments as well as the referees for fruitful comments on the manuscript.
\end{acknowledgments}



\begin{thebibliography}{99}

\bibitem{wanka90} G. Wanka, H. Hoffmann and W. Ulbricht, Colloid Polym. Sci.  268, 101 (1990) and references therein.
\bibitem{alex94} P. Alexandridis, J. F. Holzwarth and T. A. Hatton, Macromolecules 27, 2414 (1994) and references therein.
\bibitem{kozlov00} M. Y. Kozlov, N. S. Melik-Nubarov, E. V. Batrakova and A. Kabanov, Macromolecules 33, 3305 (2000) and references therein.
\bibitem{armstrong95} J. Armstrong, B. Chowdhry, R. O'Brien, A. Beezer, J. Mitchell and S. Leharne, J. Phys. Chem. 99, 4590 (1995).
\bibitem{saeki76} S. Saeki, N. Kuwahara, M. Nakata and M. Kaneko, Polymer 17, 685 (1976).
\bibitem{bae91} Y. C. Bae, S. M. Lambert and D. S. Soane, Macromolecules 24, 4403 (1991).
\bibitem{bae93} Y. C. Bae, J. J. Shim, D. S. Soane and J. M. Prausnitz, J. Appl. Polym. Sci. 47, 1193 (1993).
\bibitem{mortensen01} for a recent review see K. Mortensen, Colloids Surf. A: Physicochem. Eng. Aspects 183-185, 277 (2001).
\bibitem{almgren91} M. Almgren, J. Alsins and P. Bahadur, Langmuir 7, 446 (1991).
\bibitem{bahadur92} P. Bahadur, P. Li, M. Almgren and W. Brown, Langmuir 8, 1903 (1992).
\bibitem{alex97} P. Alexandridis and J. F. Holzwarth, Langmuir 13, 6074 (1997).
\bibitem{pedersen02} J. K. Pedersen and C. Svaneborg, Curr. Opin. Colloid Interface Sci. 7, 158 (2002).
\bibitem{brown91} W. Brown, K. Schill\'{e}n, M. Almgren, S. Hvidt and P. Bahadur, J. Phys. Chem. 95, 1850 (1991).
\bibitem{mortensen93a} K. Mortensen, J. S. Pedersen, Macromolecules 26, 805 (1993).
\bibitem{linse93} P. Linse, J. Phys. Chem. 97, 13896 (1993).
\bibitem{isra91} J. Israelachvili, Intermolecular and Surface Forces; Academic Press: San Diego, CA, 1991.
\bibitem{mortensen93b} K. Mortensen and W. Brown, Macromolecules 26, 4128 (1993).
\bibitem{schillen94} K. Schill\'{e}n, W. Brown and R. M. Johnsen, Macromolecules 27, 4825 (1994).
\bibitem{jorgensen97} E. B. J{\o}rgensen, S. Hvidt, W. Brown and K. Schill\'{e}n, Macromolecules 30, 2355 (1997).
\bibitem{liu98} Y. Liu, S.-H. Chen and J. S. Huang, Macromolecules 31, 6226 (1998).
\bibitem{michels01} B. Michels, G. Waton and R. Zana, Colloids Surf., A 183-185, 55 (2001).
\bibitem{aswal01} V. K. Aswal, P. S. Goyal, J. Kohlbrecher and P. Bahadur, Chem. Phys. Lett. 2001, 349, 458 (2001).
\bibitem{duval05} M. Duval, G. Waton and F. Schosseler, Langmuir, 21, 4904 (2005).
\bibitem{waton04} G. Waton, B. Michels, A. Steyer and F. Schosseler, Macromolecules 37, 2313 (2004).
\bibitem{mcleish86} T. C. B. McLeish and R. C. Ball, J. Polym. Sci. Polym. Phys. Ed. 24, 1735 (1986).
\bibitem{spenley93} N. A. Spenley, M. E. Cates and T. C. B. McLeish, Phys. Rev. Lett. 71, 939 (1993).
\bibitem{schmitt95} V. Schmitt, C. M. Marques and F. Lequeux, Phys. Rev. E 52, 4009 (1995).
\bibitem{spenley96} N. A. Spenley, X.-F. Yuan and M. E. Cates, J. Phys. II France 6, 551 (1996).
\bibitem{olmsted97} P. D. Olmsted and C.-Y. D. Lu, Phys. Rev. E 56, R55 (1997).
\bibitem{yuan99} X.-F. Yuan, Europhys. Lett. 46, 542 (1999).
\bibitem{olmsted99} P. D. Olmsted, Europhys. Lett. 48, 339 (1999).
\bibitem{goveas99} J. L. Goveas and D. J. Pine, Europhys. Lett. 48, 706 (1999).
\bibitem{radu00} O. Radulescu and P. D. Olmsted, J. Non-Newtonian Fluid Mech. 91, 143 (2000).
\bibitem{olmsted00} P. D. Olmsted and O. Radulescu, J. Rheol. 44, 257 (2000).
\bibitem{lu00} C.-Y. Lu, P. D. Olmsted and R. C. Ball, Phys. Rev. Lett. 84, 642 (2000).
\bibitem{goveas01} J. L. Goveas and P. D. Olmsted, Eur. Phys. J. E 6, 79 (2001).
\bibitem{yuan02} X.-F. Yuan and L. Jupp, Europhys. Lett. 60, 691 (2002).
\bibitem{bautista02} F. Bautista, J. F. A. Soltero, E. R. Mac\'{i}as and J. E. Puig, J. Phys. Chem. B 106, 13018 (2002).
\bibitem{fielding03} S. M. Fielding and P. D. Olmsted, Eur. Phys. J. E 11, 65 (2003).
\bibitem{fielding03b} S. M. Fielding an P. D. Olmsted, Phys. Rev. Lett. 90, 224501 (2003).
\bibitem{fielding03c} S. M. Fielding and P. D. Olmsted, Phys. Rev. E 68, 036313 (2003).
\bibitem{fielding04} S. M. Fielding and P. D. Olmsted, Phys. Rev. Lett. 92, 084502 (2004).
\bibitem{cook04} L. P. Cook and L. F. Rossi, J. Non-Newtonian Fluid Mech. 116, 347 (2004).
\bibitem{aradian05} A. Aradian and M. E. Cates, Europhys. Lett. 70, 397 (2005).
\bibitem{mikh05} V. S. Mikhailenko, E. E. Scime and V. V. Mikhailenko, Phys. Rev. E 71, 026306 (2005).
\bibitem{fielding05} S. M. Fielding, Phys. Rev. Lett. 95, 134501 (2005).
\bibitem{yesilata06} B. Yesilata, C. Clasen and G.H. McKinley, J. Non-Newtonian Fluid Mech. 133, 73 (2006).
\bibitem{rossi06} L. F. Rossi, G. McKinley and L. P. Cook, J. Non-Newtonian Fluid Mech. 136, 79 (2006).
\bibitem{rehage91} H. Rehage and H. Hoffmann, Mol. Phys. 74, 933 (1991).
\bibitem{khatory93} A. Khatory, F. Lequeux, F. Kern and S. J. Candau, Langmuir 9, 1456 (1993).
\bibitem{berret94} J.-F. Berret, D. C. Roux and G. Porte, J. Phys. II France 4, 1261 (1994).
\bibitem{mair96} R. W. Mair and P. T. Callaghan, Europhys. Lett. 36, 719 (1996).
\bibitem{britton97} M. M. Britton and P. T. Callaghan, Phys. Rev. Lett. 78,4930 (1997).
\bibitem{bolten97} P. Boltenhagen, Y. Hu, E. F. Matthys and D. J. Pine, Europhys. Lett. 38, 389 (1997).
\bibitem{koch98} S. Koch, T. Schneider and W. K\"uter, J. Non-Newtonian Fluid Mech. 78, 47 (1998).
\bibitem{britton99} M.M. Britton and P.T. Callaghan, Eur. Phys. J. B 7, 237 (1999).
\bibitem{fischer00} E. Fischer and P. T. Callaghan, Europhys. Lett. 50, 803 (2000).
\bibitem{fischer02} P. Fischer, E. K. Wheeler and G. G. Fuller, Rheol. Acta 41, 35 (2002).
\bibitem{escal03} J. I. Escalante, E. R. Mac\'{i}as, F. Bautista, J. H. P\'erez-L\'opez, J. F. A. Soltero and J. E. Puig, Langmuir 19, 6620 (2003).
\bibitem{salmon03} J.-B. Salmon, A. Colin, S. Manneville and F. Molino, Phys. Rev. Lett. 90, 228303 (2003).
\bibitem{mendez03} A. F. M\'endez-S\'anchez, M. R. L\'opez-Gonz\'alez, V. H. Rol\'on-Garrido, J. P\'erez-Gonz\'alez Lourdes de Vargas, Rheol. Acta 42, 56 (2003).
\bibitem{radu03} O. Radulescu, P. D. Olmsted, J. P. Decruppe, S. Lerouge, J.-F. Berret and G. Porte, Europhys. Lett. 62, 230 (2003).
\bibitem{lerouge04} S. Lerouge, J.-P. Decruppe and P. Olmsted, Langmuir 20, 11355 (2004).
\bibitem{lopez04} M. R. L\'opez-Gonz\'alez, W. M. Holmes, P. T. Callaghan and P. J. Photinos, Phys. Rev. Lett. 93, 268302 (2004).
\bibitem{becu04b} L. B\'{e}cu, S. Manneville and A. Colin, Phys. Rev. Lett. 93, 018301 (2004).
\bibitem{herle05} V. Herle, P. Fischer and E. J. Windhab, Langmuir 21, 9051 (2005).
\bibitem{lee05} J. Y. Lee, G. G. Fuller, N. E. Hudson and X.-F. Yuan, J. Rheol. 49, 537 (2005).
\bibitem{hu05} Y. T. Hu and A. Lips, J. Rheol. 49, 1001 (2005).
\bibitem{azzouzi05} H. Azzouzi, J.P. Decruppe, S. Lerouge and O. Greffier, Eur. Phys. J. E 17, 507 (2005).
\bibitem{vdg06} J. van der Gucht, M. Lemmers, W. Knoben, N. A. M. Besseling, and P. Lettinga, Phys. Rev. Lett 97, 108301 (2006).
\bibitem{manneville04} S. Manneville, L. B\'{e}cu and A. Colin, Eur. Phys. J. Appl. Phys. 28, 361 (2004).
\bibitem{mooney31} M. Mooney, J. Rheol 2, 210 (1931)
\bibitem{barnes95} H. A. Barnes, J. Non-Newtonian Fluid Mech. 56, 221 (1995).
\bibitem{leal80} G. Leal, Annu. Rev. Fluid Mech. 12, 435 (1980).
\bibitem{agarwal94} U. S. Agarwal, A. Dutta and R. A. Mashelkar, Chem. Eng. Sci. 49, 1693 (1994).
\bibitem{brochard92} F. Brochard-Wyart, P.-G. de Gennes, Langmuir 8, 3033 (1992).
\bibitem{brochard96} F. Brochard-Wyart, C. Gay and P.-G. de Gennes, Macromolecules 29, 377 (1996).
\bibitem{migler92} K. B. Migler, H. Hervet, and L. L\'{e}ger, Phys. Rev. Lett. 70, 287 (1993).
\bibitem{durliat97} E. Durliat, H. Hervet and L. L\'{e}ger, Europhys. Lett. 38, 383 (1997).
\bibitem{leger97} L. L\'{e}ger, H. Hervet, G. Massey and E. Durliat, J. Phys. Condens. Matter 9, 7719 (1997).
\bibitem{hervet03} H. Hervet and L. L\'{e}ger, C. R. Physique 4, 241 (2003).
\bibitem{larue04} I. LaRue, M. Adam, M. da Silva, S. S. Sheiko and M. Rubinstein, Macromolecules 37, 5002 (2004).
\bibitem{connel03} S. D. Connell, S. Collins, J. Fundin, Z. Yang and I. W. Hamley, Langmuir 19, 10449 (2003).
\bibitem{hamley04} I. W. Hamley, S. D. Connell and S. Collins, Macromolecules 37, 5337 (2004).
\bibitem{cates87} M. E. Cates, Macromolecules 20, 2289 (1987).
\bibitem{salmon03b} J.-B. Salmon, S. Manneville and A. Colin, Phys. Rev. E 68, 051503 (2003).
\bibitem{drummond03} C. Drummond, J. Israelachvili and P. Richetti, Phys. Rev. E 67, 066110 (2003).
\bibitem{drummond04} C. Drummond, M. In and P. Richetti, Eur. Phys. J. E 15, 159 (2004).
\bibitem{larson88} R. G. Larson, Constitutive Equations for Polymer Melts and Solutions; Butterworths: Boston, MA, 1988.
\bibitem{picard02} G. Picard, A. Ajdari, L. Bocquet and F. Lequeux, Phys. Rev. E 66, 051501 (2002).
\bibitem{muller89} S. J. Muller, R. G. Larson and E. S. G. Shaqfeh, Rheol. Acta 28, 499 (1989).
\bibitem{larson90} R. G. Larson , E. S. G. Shaqfeh and S. J. Muller, J. Fluid Mech. 218, 573 (1990).
\bibitem{shaqfeh96} E. S. G. Shaqfeh, Annu. Rev. Fluid. Mech. 28, 129 (1996).
\bibitem{oztekin96} G. H. McKinley, P. Pakdel and A. {\"O}ztekin, J. Non-Newtonian Fluid Mech. 67, 19 (1996).

\end{thebibliography}
\end{document}